\documentclass[preprint,showpacs,preprintnumbers,amsmath,amssymb]{revtex4}
\usepackage{graphicx}
\usepackage{dcolumn}
\usepackage{bm}

\begin{document}

\title{The $^8$Li($d, p$)$^9$Li Reaction and
the Astrophysical $^8$Li($n, \gamma$)$^9$Li Reaction Rate}% Force line breaks with \\

\author{Z. H. Li}\email{zhli@ciae.ac.cn}
\author{W. P. Liu}
\author{X. X. Bai}
\author{B. Guo}
\author{G. Lian}
\author{S. Q. Yan}
\author{B. X. Wang}
\author{S. Zeng}
\author{Y. Lu}
\author{J. Su}
\author{Y. S. Chen}
\author{K. S. Wu}
\author{N. C. Shu}
\affiliation{%
China Institute of Atomic Energy, P. O. Box 275(46),  Beijing 102
413, P. R. China}%
\author{T. Kajino}
\affiliation{
National Astronomical Observatory of Japan, Mitaka, Tokyo 181-8588, Japan;\\
Department of Astronomy, University of Tokyo, Bunkyo-ku, Tokyo
113-0033 Japan}

\date{\today}% It is always \today, today,

\begin{abstract}
The $^8$Li($n,\gamma$)$^9$Li reaction plays an important role in
both the r-process nucleosynthesis and the inhomogeneous big bang
models, its direct capture rates can be extracted from the
$^8$Li($d, p$)$^9$Li reaction, indirectly. We have measured the
angular distribution of the $^8$Li($d, p$)$^9$Li$_{\textrm{g.s.}}$
reaction at $E_{\textrm{c.m.}}$ = 7.8 MeV in inverse kinematics
using coincidence detection of $^{9}\textrm{Li}$ and recoil
proton, for the first time. Based on Distorted Wave Born
Approximation (DWBA) analysis, the $^{8}\textrm{Li}(d,
p)^{9}\textrm{Li}_{\textrm{g.s.}}$ cross section was determined to
be 7.9 $\pm$ 2.0 mb. The single particle spectroscopic factor,
$S_{1,3/2}$, for the ground state of $^{9}\textrm{Li}$ =
$^{8}\textrm{Li} \otimes n$ was derived to be $0.68\pm 0.14$, and
then used to calculate the direct capture cross sections for the
$^{8}\textrm{Li}(n, \gamma)^{9}\textrm{Li}_{\textrm{g.s.}}$
reaction at energies of astrophysical interest. The astrophysical
$^{8}\textrm{Li}(n, \gamma)^{9}\textrm{Li}_{\textrm{g.s.}}$
reaction rate for the direct capture was found to be 3970 $\pm$
950 $\textrm{cm}^{3}\textrm{mole}^{-1}s^{-1}$ at $T_9$ = 1. This
presents the first experimental constraint for the
$^{8}\textrm{Li}(n, \gamma)^{9}\textrm{Li}$ reaction rates of
astrophysical relevance.
\end{abstract}

\maketitle

The $^{8}\textrm{Li}(n,\gamma)^{9}\textrm{Li}$ reaction has
attracted much attention in the recent years because of its
importance in astrophysics. In the explosive neutron-rich
environments, the stability gap at mass number $A$ = 8 can be
bridged with reactions involving the unstable nucleus
$^8\textrm{Li}$ to synthesize $A$ $>$ 8 nuclides. Type II
supernovae and inhomogeneous big bang nucleosynthesis are thought
to be the astrophysical sites for such nucleosynthetic processes.

In the recently developed scenario of the r-process
nucleosynthesis occurring in the circumstances of type II
supernovae \cite{Ter01,Kaj02}, all preexisting nuclei, in the
region between the nascent neutron star and the shock front, are
believed to have been photo-disintegrated to protons and neutrons
at the initial high temperature. With the descent of temperature,
they recombine mostly under nuclear statistical equilibrium,
leading to the production of seed nuclei for r-process. When a
high $\alpha$-particle abundance forms, the main reaction chain is
initiated by $^{4}\textrm{He}(\alpha
n,\gamma)^{9}\textrm{Be}(\alpha,n)^{12}\textrm{C}$. After
$\alpha$-rich freezeout, however,
$^{4}\textrm{He}(t,\gamma)^{7}\textrm{Li}(n,\gamma)^{8}\textrm{Li}(\alpha,n)^{11}\textrm{B}$
becomes one of the active reaction chains to produce seed nuclei.
In this phase, the $^{8}\textrm{Li}(n,\gamma)^{9}\textrm{Li}$
reaction, that leads to the production of $^{9}\textrm{Be}$ via
the $^{9}\textrm{Li}$ $\beta$-decay, is in competition with the
$^{8}\textrm{Li}(\alpha, n)^{11}\textrm{B}$ reaction. The
competition determines which reaction path is taken. This reaction
also competes with the $^{8}\textrm{Li}$ $\beta$-decay. Depending
on the rate, $^{8}\textrm{Li}(n,\gamma)^{9}\textrm{Li}$ may
influence the abundances of seed nuclei to some extent. Recently,
binary neutron star mergers have been proposed as the possible
alternative sites for the r-process \cite{Ros01}, in which similar
situation is also found.

The inhomogeneous big bang models \cite{Kaj90} predict relatively
higher abundances of $A$ $>$ 8 nuclides than the standard model
does. An unconvincing agreement \cite{Hat95} between the observed
primordial abundances and predicted ones with the standard model
may be a hint for the need of inhomogeneous models, in which
$^{7}\textrm{Li}(n,\gamma)^{8}\textrm{Li}(\alpha,n)^{11}\textrm{B}$
are generally thought to be the major reaction chain forming
heavier nuclei. However, it has been found that
$^{7}\textrm{Li}(n,\gamma)^{8}\textrm{Li}(n,\gamma)^{9}\textrm{Li}(\alpha,n)^{12}\textrm{B}$
may be even more important \cite{Rau94}. Thus the
$^{8}\textrm{Li}(n,\gamma)^{9}\textrm{Li}$ reaction affects not
only the reaction path to $A$ $>$ 8 isotopes but also the
abundances of Li, Be, B and C.

Considerable effort has been devoted to experimentally determining
the $^{8}\textrm{Li}(\alpha,n)^{11}\textrm{B}$ cross section as
described in Ref. \cite{Miy04} and references therein. However, a
large uncertainty still remains for the
$^{8}\textrm{Li}(n,\gamma)^{9}\textrm{Li}$ cross section. There
were some microscopic and systematic calculations of this reaction
that deviated by order of magnitude
\cite{Rau94,Mal89,Mao91,Thi91,Des93,Ber99}. Direct measurement of
the $^{8}\textrm{Li}(n,\gamma)^{9}\textrm{Li}$ reaction is
impossible because no neutron target exists and the half life of
$^8$Li is too short (838 ms) as a target. The only experimental
information was obtained from two Coulomb dissociation
measurements \cite{Kob03,Zec98} that presented the different upper
limits. It is therefore highly needed to measure the
$^{8}\textrm{Li}(n,\gamma)^{9}\textrm{Li}$ cross section through
an independent approach. It is a practicable method to extract the
direct capture cross section for the
$^{8}\textrm{Li}(n,\gamma)^{9}\textrm{Li}$ reaction using the
direct capture model \cite{Rol73,Moh03} and spectroscopic factor
which can be deduced from the angular distribution of the transfer
reaction $^{8}\textrm{Li}(d, p)^{9}\textrm{Li}$.

 Up to now, the only measurement of the $^{8}\textrm{Li}(d,
 p)^{9}\textrm{Li}$ reaction was performed in 1990's \cite{Bal95}, in which an upper limit of cross section was
 presented though no $^{9}\textrm{Li}$ events were detected. In present work, we measured the angular
 distribution of the $^{8}\textrm{Li}(d, p)^{9}\textrm{Li}_{\textrm{g.s.}}$($Q$ = 1.839 MeV) reaction at $E_{\textrm{c.m.}} =
7.8$ MeV in inverse kinematics and derived the
$^{8}\textrm{Li}(n,\gamma)^{9}\textrm{Li}_{\textrm{g.s.}}$ direct
capture cross sections at energies of astrophysical interest.

\begin{figure}
\includegraphics[height=8.5 cm,angle=-90]{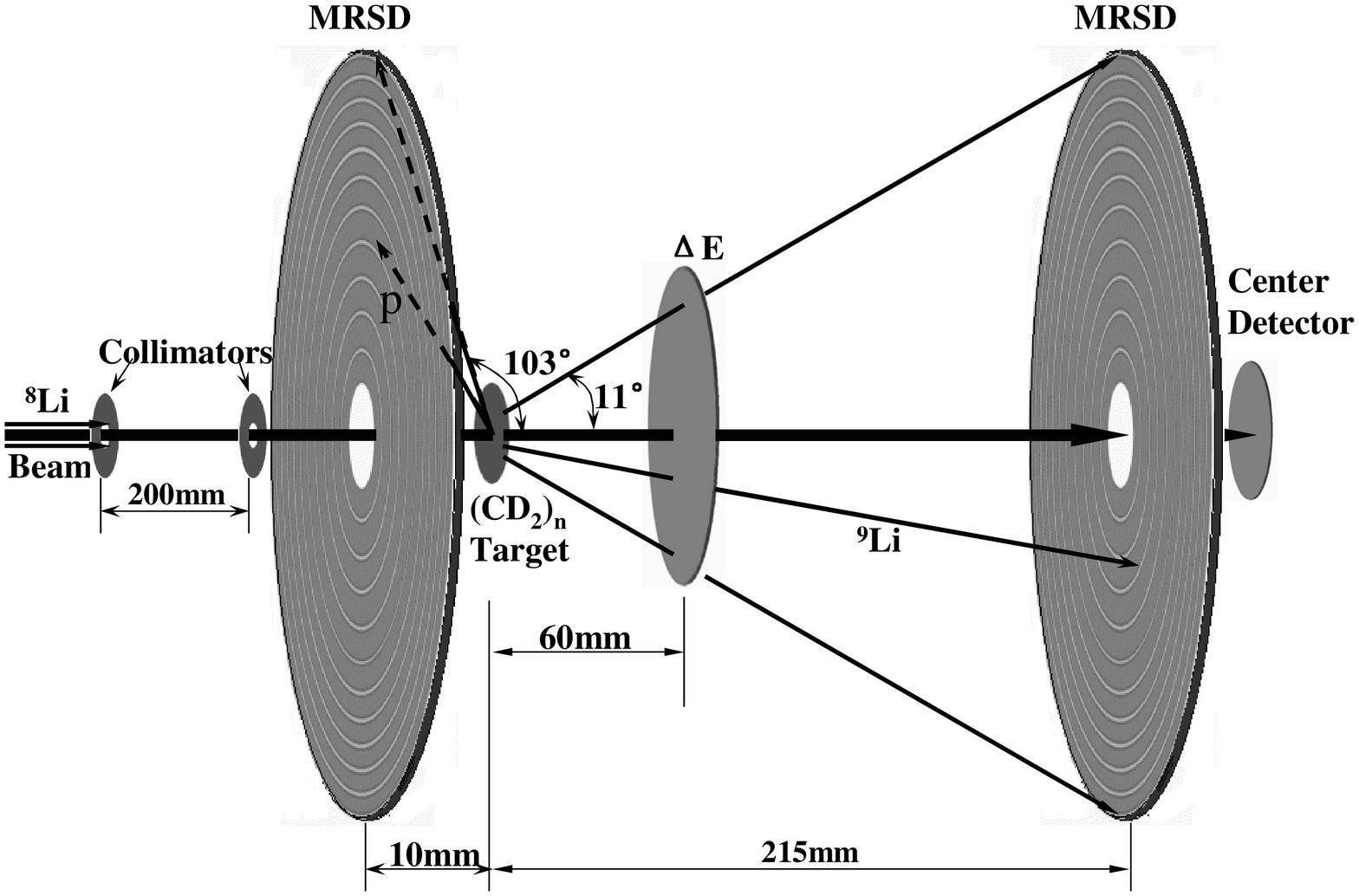}
\caption{\label{fig:setup}Schematic layout of the experimental
setup.}
\end{figure}

The experiment was carried out using the secondary beam facility
GIRAFFE \cite{Liu03} of the HI-13 tandem accelerator, Beijing. A
44 MeV $^{7}$Li primary beam from the tandem impinged on a 4.8 cm
long deuterium gas cell at a pressure of 1.6 atm. The front and
rear windows of the gas cell were Havar foils, each in thickness
of 1.9 mg/cm$^2$. The $^8$Li ions were produced via the
$^{2}\textrm{H}(^{7}\textrm{Li}, ^{8}\textrm{Li})^{1}\textrm{H}$
reaction. After the magnetic separation and focus with a dipole
and a quadrupole doublet, the 39 MeV $^{8}$Li secondary beam was
delivered. Typical purity of the $^8$Li beam was about 80\%, the
main contaminants were $^7$Li ions out of Rutherford scattering of
the primary beam in the gas cell windows as well as beam tube. The
$^{8}$Li beam was then collimated with two apertures in diameter
of 3 mm and directed onto a $(\textrm{CD}_{2})_{n}$ target in
thickness of 1.5 $\textrm{mg}/\textrm{cm}^2$ to study the
$^{2}\textrm{H}(^{8}\textrm{Li},^{9}\textrm{Li})^{1}\textrm{H}$
reaction, the typical beam intensity on the target was
approximately 1000 pps. The beam energy spread on the target was
0.52 MeV FWHM for long-term measurement, and the beam angular
divergence was about $\pm$ $0.3^{\circ}$.  A carbon target in
thickness of $1.8$ $\textrm{mg}/\textrm{cm}^2$ served as the
background measurement.

The experimental setup is shown in Fig.~\ref{fig:setup}. Two 300
$\mu$m thick Multi-Ring Semiconductor Detectors (MRSDs) with
center holes were used. The upstream one aimed at detection of the
recoil protons, and the downstream one backed by an independent
300 $\mu$m thick silicon detector placed at the center hole served
as a residue energy $E_{r}$ detector which composed a $\Delta E -
E_{\textrm{r}}$ counter telescope together with a 21.7 $\mu$m
thick silicon $\Delta E$ detector. Such a detector configuration
covered the laboratory angular ranges from $0^\circ$ to
$11^{\circ}$ (greater than the maximum $\theta_{\textrm{lab}}$ of
$^9$Li, 10.7$^{\circ}$) and from 103$^{\circ}$ to 170$^{\circ}$
respectively. For coincidence measurement, the detectable angular
range of $^9$Li in center of mass system was from 10$^\circ$ to
50$^\circ$. This setup also facilitated to precisely determine the
accumulated quantity of incident $^8$Li because the $^8$Li
themselves were recorded by the counter telescope simultaneously.

\begin{figure}
\includegraphics[height=5.0 cm]{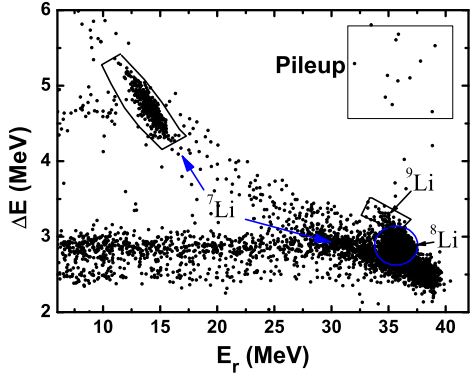}
\caption{\label{fig:eer}Scatter plot of  $\Delta E$ $vs.$ $E_{r}$
for the $^{9}\textrm{Li}$-proton coincidence measurement with
$(\textrm{CD}_{2})_{n}$ target. The 4 regions were selected to
scrutinize the random coincidence events.}
\end{figure}

\begin{figure}
\includegraphics[height=6.0 cm]{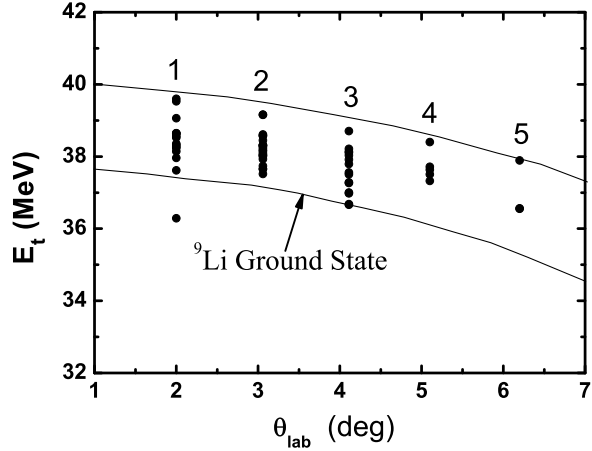}
\caption{\label{fig:etthe}Scatter plot of $E_{t}$ $vs.$
$\theta_{\textrm{lab}}$ for the events within the $^{9}$Li gate in
Fig. 2. The ring numbers of the downstream MASD are indicated.}
\end{figure}

The accumulated quantity of incident $^8$Li for the
$(\textrm{CD}_{2})_{n}$ target measurement was approximately
$1.66\times 10^8$, and 1.11 $\times$ $10^{7}$ for measurement with
the carbon target. Fig.~\ref{fig:eer} displays the summing $\Delta
E-E_{r}$ scatter plot of the coincidence events over all the rings
of downstream MRSD. In order to scrutinize the origin of
non-$^9$Li region events in Fig.~\ref{fig:eer}, four typical
regions were selected in which the ratios of events in coincidence
spectrum to those in non-coincidence one were compared. It was
found that the ratio in the $^{9}$Li region is much higher than
those (nearly constant) in other three regions. We can thus
conclude that the events in non-$^{9}$Li regions result from the
random coincidence caused by the noise tail in the proton spectrum
of the upstream MRSD. The scatter plot of $E_{t}$ vs. $\theta
_{\textrm{lab}}$ for the events within $^{9}$Li gate in
Fig.~\ref{fig:eer} is shown in Fig.~\ref{fig:etthe}. The zone
between two solid lines stands for the kinematics region of
$^{9}$Li ground state, based on Monte Carlo simulation. This
figure further demonstrates that the events within $^{9}$Li gate
in Fig.~\ref{fig:eer} are the true $^{9}$Li products. Actually,
the selection of $^9$Li events was individually performed for each
ring of downstream MRSD. The energies and angles of protons on
upstream MRSD relevant to the $^9\textrm{Li}$ events appearing on
the specific ring of downstream MRSD were restricted with the
kinematics. Thus, the random coincidence events were effectively
depressed by setting the corresponding windows for energies and
angles (ring numbers) of protons. Finally about 50 $^9$Li events
were identified for the measurement with $(\textrm{CD}_{2})_{n}$
target and no background event was found in the same region for
the carbon target run.

\begin{figure}
\includegraphics[height=5.8 cm]{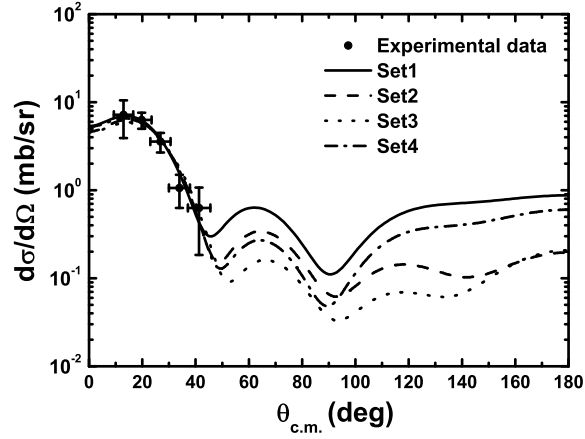}
\caption{\label{fig:andis}Measured angular distribution of
$^{8}\textrm{Li} (d, p)^{9}\textrm{Li}_{\textrm{g.s.}}$ at
$E_{\textrm{c.m.}}$ = 7.8 MeV, together with DWBA calculations
using different optical potential parameters.}
\end{figure}

Because of the existence of dead gaps in both MRSDs, a Monte Carlo
simulation was used to calculate the coincidence efficiency
between $^{9}$Li ions detected by downstream MRSD and protons by
upstream one. The simulation took the geometrical factor, angular
straggling and energy straggling into account. The coincidence
efficiency was deduced from the ratio of the proton events in the
relevant rings of upstream MRSD to the $^9$Li events in the
specific ring of the downstream one. The differential cross
sections can then be deduced. The resultant angular distribution
is shown in Fig.~\ref{fig:andis}. The uncertainties of
differential cross sections mainly arise from the statistics and
the assignment of $^{9}$Li gate; the angular uncertainties are
from the angular divergence of $^{8}$Li beam, the finite size of
beam spot, the angular straggling in the target and $\Delta E$
detector as well as the width of each ring of the downstream MRSD.

The spin and parity of $^{8}$Li and $^{9}$Li (ground state) are
$2^{+}$ and $3/2^{-}$, respectively. The $^{8}\textrm{Li}(d,
p)^{9}\textrm{Li}_{\textrm{g.s.}}$ cross section involves two
components corresponding to $(l = 1, j = 3/2)$ and $(l = 1, j =
1/2)$ transfers, the differential cross section can be expressed
as
%\begin{widetext}
\begin{equation}\label{eq1}
({d\sigma \over d\Omega})_{exp} =
S_{d}[S_{1,3/2}\sigma_{1,3/2}(\theta)+S_{1,1/2}\sigma_{1,1/2}(\theta)],
\end{equation}
%\end{widetext}
where $({d\sigma \over d\Omega})_{\textrm{exp}}$ and
$\sigma_{l,j}(\theta)$ denote the measured and DWBA differential
cross sections respectively, $S_{1,3/2}$  and $S_{1,1/2}$ are
spectroscopic factors for the ground state of $^9$Li = $^8$Li
$\otimes n$, corresponding to the $j =3/2$ and $1/2$ orbits. $S_d$
is the spectroscopic factor of deuteron that is close to unity.
The $S_{1,3/2}$ and $S_{1,1/2}$ can be extracted by normalizing
DWBA differential cross sections to the experimental data.

The finite-range DWBA code PTOLEMY \cite{Mac78} was used to
compute the angular distribution. In the calculation, only the
neutron transfer to the $1p3/2$ orbit was taken into account
because the contribution of the $1p1/2$ orbit is less than 5\%
\cite{Moh03,Bea01}. All the entrance channel parameters were taken
from Ref. \cite{Per76}, the exit channel ones from Refs.
\cite{Per76} and \cite{Wat69}, respectively. Fig.~\ref{fig:andis}
presents the normalized angular distributions for four sets of
optical potential parameters, each curve corresponds to one
spectroscopic factor, $S_{1,3/2}$. The average value of these
spectroscopic factors was found to be 0.68 $\pm$ 0.14 with the
``standard'' bound state potential parameters (radius $r_0$ = 1.25
fm, diffuseness $a$ = 0.65 fm), it fairly agrees with the result
$S_{1,3/2}$ = 0.73 extracted from the mirror reaction
$^{8}\textrm{B}(d, n)^{9}\textrm{C}$ \cite{Bea01}. The cross
section for the $^8\textrm{Li}(d,
p)^{9}\textrm{Li}_{\textrm{g.s.}}$ reaction at $E_{\textrm{c.m.}}$
= 7.8 MeV was determined to be 7.9 $\pm$ 2.0 mb through
integration of the calculated angular distributions. The
uncertainties of the spectroscopic factor and cross section result
mainly from the difference of the optical potentials used in the
calculation, as well as the statistical error of the measurement.

\begin{figure}
\includegraphics[height=5.5 cm]{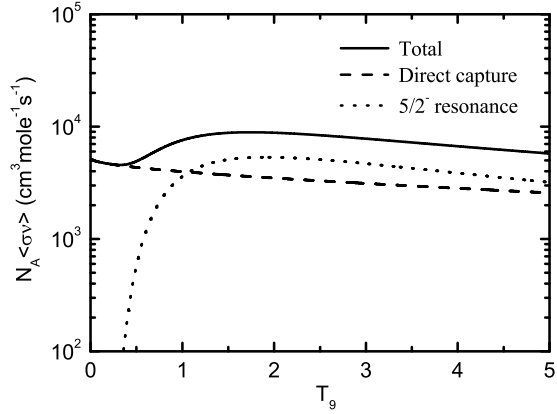}
\caption{\label{fig:ngcross}Temperature dependence of the reaction
rates for $^{8}\textrm{Li}(n,\gamma)^{9}\textrm{Li}$. The
contribution of direct capture is the result of present work and
that of 5/2$^-$ resonance is taken from of Ref. \cite{Rau94}.}
\end{figure}

\begin{figure}
\includegraphics[height=5.0 cm]{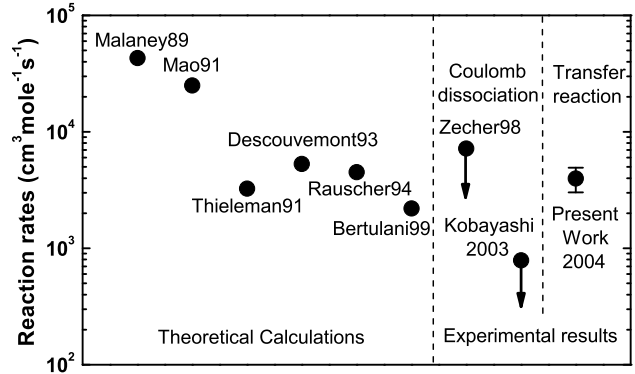}
\caption{\label{fig:rate}The
$^{8}\textrm{Li}(n,\gamma)^{9}\textrm{Li}$ reaction rates for the
direct capture derived from theoretical calculations and
experiments.}
\end{figure}

The $^{8}\textrm{Li}(n, \gamma)^{9}\textrm{Li}_{\textrm{g.s.}}$
cross section was calculated by assuming that the reaction
proceeds by direct E1 capture of neutron to the ground state of
$^9$Li.  At low energies of astrophysical interest, the
contribution of \textit{d}-wave is negligible, the capture
reaction is almost totally determined by the \textit{s}-wave
neutron capture process. The cross section for E1 capture of
neutron to the ground state of $^{9}$Li with the orbital and total
angular momenta $l_{f}$ and $j_{f}$  is given by
%\begin{widetext}
\begin{eqnarray}\label{eq2}
\sigma_{(n, \gamma)} = {16 \pi \over 9}({E_{\gamma} \over \hbar
c})^{3}{e_{eff}^{2} \over k^{2} }{1 \over \hbar v}{(2j_{f}+1)
\over
     (2I_{1}+1)(2I_{2}+1) }S_{l_{f}j_{f}}\nonumber\\
 \times|\int^{\infty}_{0}r^{2} w_{l_i}(kr)u_{l_f}(r)dr|^{2} ,
\end{eqnarray}
%\end{widetext}
where $E_{\gamma}$ stands for the $\gamma$-ray energy, $v$ is the
relative velocity between particles 1 (neutron) and 2 ($^{8}$Li),
$k$ the incident wave number, $I_{i}$ the spin of particle $i$.
$e_{\textrm{eff}}$ = $-eZ/(A+1)$ is the neutron effective charge
for the E1 transition in the potential produced by a target
nucleus with mass number $A$ and atomic number $Z$. $w_{l_i}(kr)$
is the distorted radial wave function for the entrance channel,
$u_{l_f}(r)$ the radial wave function of the bound state neutron
in $^9$Li which can be calculated by solving the respective
Schr\"{o}dinger equation. The optical potential for the neutron
scattering on unstable nucleus $^8$Li is unknown experimentally.
We adopted a real folding potential which was calculated using the
$^8$Li density distribution from the measured interaction cross
section \cite{Tan85} and an effective nucleon-nucleon interaction
DDM3Y \cite{Kob84}. The imaginary part of the potential is very
small because of the small flux into other reaction channels and
can be neglected in most cases involving neutron capture reaction.
The depth of the real potential was scaled to $J_V/A$ = 743 MeV
fm$^3$, the volume integral of potential per nucleon. Usually, the
optical potential changes considerably for different nuclei,
whereas the volume integral of potential per nucleon is a more
stable quantity relatively. First, we calculated the folding
potentials for $^{6,7}$Li and $^{12}$C, whose neutron capture
cross sections were measured \cite{Ohs00,Bla96,Nag91}, and found
$J_V/A$ = 749 $\pm$ 23, 729 $\pm$ 75 and 742 $\pm$ 12 MeV fm$^3$,
respectively, by fitting to the experimental cross sections. One
can see that these values are close to each other, indicating the
stability of the potential volume integral. Then the weighted
average of above values was taken as the $J_V/A$ value for $^8$Li.
As soon as the potential is known, the capture cross sections can
be calculated with the Eq. (\ref{eq2}). With the spectroscopic
factor extracted from the present experiment, the energy
dependence of direct capture cross sections for the
$^{8}\textrm{Li}(n,\gamma)^{9}\textrm{Li}_{\textrm{g.s.}}$
reaction were calculated, showing a deviation from the usual $1/v$
behavior. Fig. \ref{fig:ngcross} demonstrates the temperature
dependence of the reaction rate for the direct capture in
$^8\textrm{Li}(n,\gamma)^9\textrm{Li}$ together with that for the
resonant capture at 5/2$^-$ 4.3 MeV state in $^9$Li (deduced by
Rauscher et al. \cite{Rau94} with $\Gamma_{\gamma}$ = 0.65 eV and
$\Gamma_{\textrm{tot}}$ = 100 keV), as well as the total reaction
rate. Direct capture to the first excited state at 2.69 MeV is not
include in the calculation since the transition strength is
believed to be negligible \cite{Kob03,Zec98}. It can clearly be
seen that the direct capture plays an important role in the
$^8\textrm{Li}(n,\gamma)^9\textrm{Li}$ reaction especially in the
astrophysical environment of $T_9$ $<$ 1. The reaction rate for
the direct capture was found to be $N_{A}\langle\sigma v\rangle$ =
3970 $\pm$ 950 $\textrm{cm}^{3}\textrm{mole}^{-1}s^{-1}$ at $T_9$
= 1, the uncertainty arises from the errors of spectroscopic
factor and the volume integral of potential per nucleon. The
$^{8}\textrm{Li}(n,\gamma)^{9}\textrm{Li}$ reaction rates for the
direct capture derived from theoretical calculations and
experiments are shown in Fig. \ref{fig:rate}, our result is
significantly higher than the upper limit from the most recent
Coulomb dissociation experiment \cite{Kob03} and approximately in
agreement with the theoretical estimations reported in Refs.
\cite{Rau94,Thi91}.

In summary, we have measured the angular distribution of the
$^{8}\textrm{Li}(d, p)^{9}\textrm{Li}_{\textrm{g.s.}}$ reaction at
$E_{\textrm{c.m.}}$ = 7.8 MeV, through coincidence detection of
$^{9}$Li and recoil proton, and obtained the cross section. By
using the spectroscopic factor deduced from the
$^{8}\textrm{Li}(d, p)^{9}\textrm{Li}_{\textrm{g.s.}}$ angular
distribution, we have successfully derived the
$^{8}\textrm{Li}(n,\gamma)^{9}\textrm{Li}_{\textrm{g.s.}}$ direct
capture cross section and
astrophysical reaction rate for the first time.\\

%\begin{acknowledgments}
This work was supported by the Major State Basic Research
Development Program under Grant Nos. G200077400 and 2003CB716704,
the National Natural Science Foundation of China under Grant Nos.
10375096, 10025524 and 19935030.
%\end{acknowledgments}
%\newpage %Just because of unusual number of tables stacked at end
\bibliography{li8dp}% Produces the bibliography via BibTeX.
\end{document}